\documentclass[conference]{IEEEtran}
\IEEEoverridecommandlockouts
\usepackage{amsmath,amssymb,mathtools}
\usepackage{booktabs}
\usepackage{multirow}
\usepackage{graphicx}
\usepackage[caption=false,font=footnotesize]{subfig}
\usepackage{cite}
\usepackage{xcolor}
\usepackage{url}
\usepackage{placeins}
\usepackage{subfig}
\newcommand{\cC}{\mathcal{C}}
\newcommand{\cF}{\mathcal{F}}
\newcommand{\bphi}{\boldsymbol{\phi}}
\newcommand{\bh}{\mathbf{h}}
\newcommand{\bg}{\mathbf{g}}
\newcommand{\bw}{\mathbf{w}}
\newcommand{\bG}{\mathbf{G}}
\newcommand{\bPhi}{\boldsymbol{\Phi}}
\begin{document}

\title{\textcolor{black}{Generative AI-Enabled Mission-Aware Radio Orchestration for RIS-Assisted LEO Satellite ISAC Systems}}

\author{\IEEEauthorblockN{Fitsum Debebe Tilahun %\IEEEauthorrefmark{1},
 and
Chung G. Kang}  % \IEEEauthorrefmark{1}
\IEEEauthorblockA{School of Electrical Engineering, Korea University, Seoul, Republic of Korea\\
Email: \{fitsum\_debebe, ccgkang\}@korea.ac.kr
}}   % \IEEEauthorrefmark{1}  \IEEEauthorrefmark{1}

\maketitle

\begin{abstract}
\color{black}
{\color{black}Mission-adaptive low-Earth-orbit (LEO) satellite networks with integrated sensing and communication (ISAC) must retarget radio resources as operator goals change. To enable this adaptation from flexible operator language, we develop a generative-AI-enabled radio-orchestration framework in which a large language model (LLM) maps each mission into a structured policy comprising communication, sensing, and fairness weights, mandatory quality-of-service thresholds, power-allocation guidance, and solver initialization. Deterministic validation and physical-layer optimization then enforce feasibility and realize the policy through beam, power, and reconfigurable intelligent surface (RIS) configuration.} This mixed-timescale design uses generative AI for semantic adaptation at the mission timescale while retaining conventional wireless optimization at the faster channel timescale. We compare zero-shot (LLM-ZS) and in-context (LLM-ICL) operation on familiar and held-out compositional missions. On held-out instructions, LLM-ZS and LLM-ICL achieve $91.7\%$ and $94.4\%$ priority-order accuracy, respectively, with ICL mainly improving numerical calibration. Their downstream radio-performance difference is statistically unresolved because both usually recover the hard constraints that determine admissible actions. Accordingly, LLM-ZS is the low-context default, while LLM-ICL is useful for semantically difficult missions requiring finer calibration. Explicit alternating optimization preserves the qualitative ordering when active beams and RIS phases are optimized directly. The results show how generative AI can enhance next-generation radio orchestration without replacing feasibility-critical physical-layer optimization.
\color{black}
\end{abstract}


\section{Introduction}
Low-Earth-orbit (LEO) non-terrestrial networks can extend wireless coverage to remote and infrastructure-limited areas, but their physical-layer operation is challenged by large propagation loss, time-varying geometry, and frequent beam reconfiguration~\cite{3gpp,cellfree}. Reconfigurable intelligent surfaces (RISs) can mitigate these limitations by passively shaping the propagation environment~\cite{wuzhang,zhengleo}, while integrated sensing and communication (ISAC) allows the same spectrum and transmit hardware to support both data delivery and target sensing~\cite{liu_isac,xu_isac,luo_isac}. Combining these technologies creates a coupled radio-design problem involving communication beamforming, a sensing waveform, the communication-sensing power split, and RIS phases.

For a fixed channel realization, multiple feasible radio configurations may emphasize different combinations of user throughput, target sensing, and weakest-user protection. The preferred operating point is therefore determined jointly by the wireless state and the mission. For example, an operator may request maximum aggregate throughput, stronger target sensing, protection of the weakest receiver, or sensing priority subject to a minimum communication requirement. These objectives require the available transmit power and RIS-assisted gain to be distributed differently among the user links, target illumination, and fairness protection. Hence, changing the mission can require a different radio configuration even when the channel itself is unchanged.

{\color{black}
Mission intent is commonly expressed in natural language rather than through optimization variables. A single instruction may combine objectives, order their importance, negate one priority, or distinguish a preferred objective from a mandatory requirement. This motivates generative AI for adaptive network control: an LLM can synthesize structured priorities and constraints for unseen objective combinations without expanding a fixed taxonomy. We therefore use the LLM as a semantic orchestration layer rather than as a high-dimensional radio optimizer. It generates mission weights, QoS thresholds, and solver guidance; deterministic validation and radio optimization then realize the policy through power allocation, beamforming, and RIS control. This exploits generative AI for flexible intent synthesis while retaining numerical optimization for feasibility-critical execution, consistent with tool-coupled AI~\cite{zhang_agents} and distinct from direct LLM-based numerical optimization~\cite{noh_llm}. Because mission intent changes more slowly than the channel, the validated policy is cached while the physical layer updates the radio variables.
}

{\color{black}
Accordingly, our contributions are threefold. First, we formulate generative-AI-enabled mission adaptation as structured policy generation over continuous objective weights, mandatory QoS thresholds, and low-dimensional search guidance, allowing unseen combinations of communication, sensing, and fairness priorities. Second, we develop a mixed-timescale architecture with zero-shot and in-context modes, consensus aggregation, deterministic validation, policy caching, and solver-based radio execution, so that semantic adaptation is separated from fast channel control. Third, we evaluate the chain from held-out compositional language to wireless performance using paired channels and language, policy, and physical-layer baselines, followed by explicit alternating optimization of the active beams, power split, and RIS phases. The study identifies when contextual examples improve calibration and whether those improvements alter radio-resource decisions.
}

\section{System Model and Mission-Aware Formulation}
\subsection{Network Geometry and Communication Model}
We consider a LEO satellite equipped with an $M$-element uniform linear array (ULA), serving $K$ representative single-antenna ground receivers~\cite{cellfree} while illuminating one sensing target. A fixed $N$-element RIS assists both the user and target links. Each representative receiver denotes a compact local group with approximately common large-scale geometry and statistical CSI.

Let $h$ denote the satellite altitude, $R_E$ the Earth radius, and $\epsilon_k$ the elevation angle of receiver $k$. The satellite-to-user slant distance is
\begin{equation}
 d_k=\sqrt{(R_E+h)^2-R_E^2\cos^2\epsilon_k}-R_E\sin\epsilon_k.
 \label{eq:slant}
\end{equation}
The distance $d_k$ increases as the elevation angle decreases and determines the free-space spreading loss. Accordingly, the dimensionless free-space path-loss factor is $L_{\mathrm{fs},k}=(4\pi d_k f_c/c_0)^2$, where $f_c$ and $c_0$ denote the carrier frequency and speed of light, respectively. The complete large-scale link gain additionally includes the antenna gains, atmospheric attenuation, receiver noise figure, and shadowing.

Let $\bG\in\mathbb{C}^{N\times M}$ denote the satellite-to-RIS channel, while $\bh_{d,k}\in\mathbb{C}^{M}$ and $\bh_{r,k}\in\mathbb{C}^{N}$ denote the direct satellite-to-user and RIS-to-user channels, respectively. The RIS applies the diagonal reflection response
\begin{equation}
 \bPhi=\operatorname{diag}(\bphi),
 \label{eq:ris}
\end{equation}
where $\bphi=[e^{j\theta_1},\ldots,e^{j\theta_N}]^T$ contains the element-wise reflection coefficients, $\theta_n$ is the phase shift of element $n$, and $|\phi_n|=1$ enforces passive unit-modulus reflection. Combining the direct and reflected paths, the effective channel of user $k$ is
\begin{equation}
 \bh_k^{H}(\bphi)=\bh_{d,k}^{H}+\bh_{r,k}^{H}\bPhi\bG.
 \label{eq:effective_channel}
\end{equation}
Because the same phase vector $\bphi$ appears in every effective channel, one RIS configuration jointly affects all users.

The satellite transmits $K$ communication streams and one sensing stream as
\begin{equation}
 \mathbf{x}=\sum_{k=1}^{K}\bw_k s_k+\bw_s s_s,
 \label{eq:tx_signal}
\end{equation}
where $\bw_k\in\mathbb{C}^{M}$ and $\bw_s\in\mathbb{C}^{M}$ are the beamformers for user $k$ and the sensing waveform, respectively, while $s_k$ and $s_s$ have unit average power. The received signal at user $k$ is $y_k=\bh_k^H\mathbf{x}+n_k$, where $n_k$ has variance $\sigma_k^2$. Treating the remaining communication streams and sensing-waveform leakage as interference, with receiver noise also included, the SINR of user $k$ is given by
\begin{equation}
 \gamma_k=\frac{|\bh_k^H\bw_k|^2}{\displaystyle\sum_{j\neq k}|\bh_k^H\bw_j|^2+|\bh_k^H\bw_s|^2+\sigma_k^2}.
 \label{eq:sinr}
\end{equation}
Accordingly, the achievable spectral efficiency of user $k$ is $R_k=\log_2(1+\gamma_k)$ bit/s/Hz. We then define $R_{\Sigma}=\sum_{k=1}^{K}R_k$ as the aggregate throughput and $R_{\mathrm{wu}}=\min_k R_k$ as the weakest-user rate.

\subsection{Sensing Model}
Let $\bg_d\in\mathbb{C}^{M}$ and $\bg_r\in\mathbb{C}^{N}$ denote the direct satellite-to-target and RIS-to-target channels, respectively. Following the communication model, the effective target channel combines the direct and RIS-assisted paths as
\begin{equation}
 \bg^{H}(\bphi)=\bg_d^{H}+\bg_r^{H}\bPhi\bG.
 \label{eq:target_channel}
\end{equation}
The nominal target-return power follows the monostatic radar relation $P_{r,s}=P_tG_tG_r\sigma_{\mathrm{rcs}}\lambda_c^2/[(4\pi)^3d_t^4L_s]$, where $P_t$ is the transmit power, $G_t$ and $G_r$ are the antenna gains, $\sigma_{\mathrm{rcs}}$ is the target radar cross section, $\lambda_c$ is the wavelength, $d_t$ is the target range, and $L_s$ collects additional losses. In particular, the $d_t^{-4}$ dependence captures two-way propagation. Since both the communication beams and the dedicated sensing beam illuminate the target, the resulting sensing SNR is
\begin{equation}
 \Gamma_s=\frac{\displaystyle\sum_{k=1}^{K}|\bg^H\bw_k|^2+|\bg^H\bw_s|^2}{\sigma_s^2},
 \label{eq:sensing_snr}
\end{equation}
where $\sigma_s^2$ denotes the sensing-noise power. Hence, communication and sensing are coupled through the shared transmit-power budget and the target illumination contributed by the communication beams. To quantify target-angle estimation, let $\mathbf{a}(\vartheta)$ denote the ULA steering vector at target angle $\vartheta$, $\dot{\mathbf{a}}(\vartheta)$ its derivative, and $\mathbf{P}_{\mathbf{a}}^{\perp}=\mathbf{I}-\mathbf{a}\mathbf{a}^H/\|\mathbf{a}\|_2^2$ the projector onto the subspace orthogonal to $\mathbf{a}$. For $L$ sensing snapshots, the angle Cram\'er--Rao bound (CRB) is
\begin{equation}
 \mathrm{CRB}_{\vartheta}=\left[2L\Gamma_s\left\|\mathbf{P}_{\mathbf{a}}^{\perp}\dot{\mathbf{a}}(\vartheta)\right\|_2^2\right]^{-1}.
 \label{eq:crb}
\end{equation}
Thus, the bound decreases with the sensing SNR, the number of snapshots, and the angular sensitivity of the array. We consider one dominant target and neglect clutter.

\subsection{Mission-Aware Policy and Optimization Problem}
A natural-language mission is compiled into the radio-control policy vector $\mathbf{z}=[\lambda_R,\lambda_S,\lambda_F,R_{\mathrm{th}},\Gamma_{\mathrm{th}},\rho_0,m_{\mathrm{init}}]$, where $\lambda_R$, $\lambda_S$, and $\lambda_F$ are nonnegative weights for aggregate communication rate, sensing performance, and weakest-user fairness, respectively, with $\lambda_R+\lambda_S+\lambda_F=1$. The quantities $R_{\mathrm{th}}$ and $\Gamma_{\mathrm{th}}$ specify the minimum user-rate and sensing-SNR requirements, while $\rho_0$ provides an initial communication-power fraction and $m_{\mathrm{init}}$ selects a communication-oriented, sensing-oriented, or balanced solver initialization. Because the three performance measures have different units and ranges, they are mapped to bounded utilities before combination. More specifically, the normalized sum-rate utility is $\widetilde{R}_{\Sigma}=\tanh(R_{\Sigma}/a_R)$, where $a_R$ sets the saturation scale; similarly, the normalized sensing utility is $\widetilde{\Gamma}_{s}=\{1+\exp[-(\Gamma_{s,\mathrm{dB}}-b_S)/a_S]\}^{-1}$, where $a_S$ and $b_S$ control its slope and midpoint. Finally, weakest-user fairness is represented by $\widetilde{R}_{\mathrm{wu}}=\operatorname{clip}(K R_{\mathrm{wu}}/R_{\Sigma},0,1)$, which approaches one when the weakest-user rate is close to the per-user average.

Let $\mathbf{x}=\left(\{\bw_k\}_{k=1}^{K},\bw_s,\bphi,\rho\right)$ collect the physical-layer control variables, where $\rho\in[0,1]$ denotes the communication-power fraction, while $\mathbf{z}$ collects the mission priorities, QoS requirements, and solver guidance. Using these two variable groups and the normalized utilities above, the mission-weighted radio utility is expressed as
\begin{equation}
 \cF(\mathbf{x};\mathbf{z})=\lambda_R\widetilde{R}_{\Sigma}+\lambda_S\widetilde{\Gamma}_{s}+\lambda_F\widetilde{R}_{\mathrm{wu}}-\mu_EE(\mathbf{x})-\mu_{\Phi}C_{\Phi}(\bphi).
 \label{eq:mission_score}
\end{equation}
The first three terms reward communication throughput, sensing quality, and weakest-user fairness according to the compiled mission weights. In addition, $E(\mathbf{x})$ and $C_{\Phi}(\bphi)$ penalize energy consumption and RIS reconfiguration, while $\mu_E$ and $\mu_{\Phi}$ determine their relative importance.

Accordingly, the physical-layer controller selects the beamformers, power split, and RIS coefficients by solving
\begin{subequations}\label{eq:main_problem}
\begin{align}
 \max_{\{\bw_k\},\bw_s,\bphi,\rho}\quad &\cF(\mathbf{x};\mathbf{z})\label{eq:main_obj}\\
 \text{s.t.}\quad &R_k\geq R_{\mathrm{th}},\quad k=1,\ldots,K,\label{eq:rate_const}\\
 &\Gamma_s\geq \Gamma_{\mathrm{th}},\label{eq:sense_const}\\
 &\sum_{k=1}^{K}\|\bw_k\|_2^2\leq \rho P_{\max},\label{eq:comm_power}\\
 &\|\bw_s\|_2^2\leq (1-\rho)P_{\max},\label{eq:sense_power}\\
 &0\leq\rho\leq1,\label{eq:rho_const}\\
 &|\phi_n|=1,\quad n=1,\ldots,N.\label{eq:unit_const}
\end{align}
\end{subequations}
Constraints~\eqref{eq:rate_const}--\eqref{eq:sense_const} enforce the compiled QoS requirements, while~\eqref{eq:comm_power}--\eqref{eq:rho_const} allocate the transmit power according to $\rho$ and~\eqref{eq:unit_const} imposes unit-modulus RIS phases. Multiuser interference, shared sensing illumination, and common RIS coefficients make the problem non-convex. The LLM specifies the mission policy $\mathbf{z}$, while a deterministic radio solver computes the physical variables in $\mathbf{x}$ for the current channel state.

\section{\textcolor{black}{Generative AI Mission Compilation and Policy-Conditioned Radio Orchestration}}
\subsection{Mixed-Timescale Architecture}
\begin{figure}[!t]
\centering
\includegraphics[width=0.90\columnwidth]{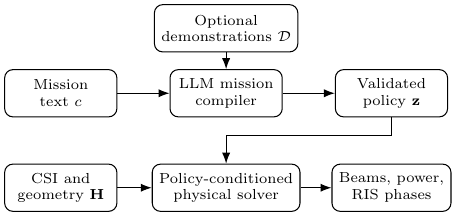}
\caption{\textcolor{black}{Generative-AI mission compilation and deterministic radio orchestration.}}
\label{fig:architecture}
\end{figure}

{\color{black}
Mission instructions typically change more slowly than satellite geometry, CSI, and interference. At the slower mission timescale, the LLM converts the operator instruction $c$ into priorities, mandatory requirements, and solver guidance; the validator checks and caches the resulting policy. At the faster channel timescale, the deterministic physical-layer optimizer maps $(\mathbf{z},\mathbf{H})$ to a feasible action $\mathbf{x}^{\star}$ by updating the beams, communication-sensing power split, and RIS phases. The LLM is invoked again only when the mission changes. Thus, generative AI adapts what the network should prioritize, while the radio solver determines how that policy is physically realized. The mixed-timescale architecture is summarized in Fig.~\ref{fig:architecture}.
}

\begin{table*}[!t]
\caption{Main simulation and implementation parameters}
\label{tab:params}
\centering
\scriptsize
\setlength{\tabcolsep}{3.0pt}
\renewcommand{\arraystretch}{0.86}
\begin{tabular}{p{0.17\textwidth}p{0.30\textwidth}p{0.17\textwidth}p{0.30\textwidth}}
\toprule
\textbf{Parameter} & \textbf{Setting} & \textbf{Parameter} & \textbf{Setting}\\
\midrule
\multicolumn{2}{l}{\textit{System, propagation, and sensing}} &
\multicolumn{2}{l}{\textit{Language experiment and evaluation}}\\
Orbit and carrier & $h=600$ km; $f_c=20$ GHz; $B=20$ MHz & Mission prompts & 12 familiar; 24 held-out\\
Arrays and users & $M=8$; $K=3$ & LLM implementation & DeepSeek-V4-Flash via DeepInfra\\
RIS configuration & primary $N=32$; sweep $N\in\{8,16,32,64\}$ & Sampling temperature & $0.05$\\
Power and gains & $P_t=10$ dBW; satellite/user gains $35/0$ dBi & Calls and demonstrations & 3 calls/prompt; 3 ICL demonstrations\\
Channel conditions & loss $1.5$ dB; shadow margin $2$ dB; noise figure $5$ dB & Monte Carlo realizations & 100 per prompt\\
Geometry and target & user $25^{\circ}$--$70^{\circ}$; target $30^{\circ}$--$65^{\circ}$; RCS $5$ dBsm & Candidate starts & 6\\
Utility scales & $a_R=6$; $a_S=4$; $b_S=1$ dB & Refinement iterations & 5\\
Penalty weights & $\mu_E=0.05$; $\mu_{\Phi}=0.04$ & Explicit validation & $N=16$; 4 phase levels; 1 coordinate pass\\
Energy proxy & $E(\mathbf{x})=0.45+0.55\rho$ & Validation trials & 6 missions; 10 trials; gain $88$ dB; $L=64$\\
\bottomrule
\end{tabular}
\end{table*}

Let $\mathcal{D}$ denote an optional set of example mission-policy pairs, referred to as demonstrations. The prompt is
\begin{equation}
 \mathcal{P}(c,\mathcal{D})=[\mathcal{R},\mathcal{S},\mathcal{D},c],
 \label{eq:prompt}
\end{equation}
where $\mathcal{R}$ is a field-definition guide, $\mathcal{S}$ specifies the structured radio-policy format, and $c$ is the current mission instruction. From this prompt, the LLM maps mandatory service requirements to $R_{\mathrm{th}}$ and $\Gamma_{\mathrm{th}}$, converts the relative importance of communication, sensing, and weakest-user fairness into $(\lambda_R,\lambda_S,\lambda_F)$, and provides search guidance through $\rho_0$ and $m_{\mathrm{init}}$. For example, ``sensing is primary, but preserve moderate communication service'' should assign the largest sensing weight, retain a nonzero rate requirement, and select a sensing-oriented initialization. Conversely, ``protect the weakest terminal and devote little effort to sensing'' should emphasize fairness and the user-rate requirement while reducing the sensing priority. The resulting proposal is
\begin{equation}
 \widehat{\mathbf{z}}=\mathcal{F}_{\mathrm{LLM}}\!\left(\mathcal{P}(c,\mathcal{D})\right),
 \label{eq:llm_output}
\end{equation}
where $\widehat{\mathbf{z}}$ is an unvalidated policy: its thresholds determine admissibility, its weights rank admissible configurations, and its guidance fields initialize the search.

\subsection{Zero-Shot and In-Context Learning Modes}
The same frozen LLM is evaluated in two prompting modes, which differ only in whether example mission-policy pairs are included in $\mathcal{D}$.

\noindent\textit{1) Zero-shot mode (LLM-ZS):} In this mode, no task-specific example is provided, so $\mathcal{D}=\varnothing$. The prompt contains only the rubric, output schema, and new mission instruction. The model therefore relies on its pretrained language knowledge together with the explicit definitions of the policy fields. Multiple responses are generated and aggregated, but no demonstration is supplied and no model parameter is updated.

\noindent\textit{2) In-context learning mode (LLM-ICL):} In this mode, the prompt additionally contains $D$ example pairs, $\mathcal{D}=\{(c_i,\mathbf{z}_i)\}_{i=1}^{D}$, where $c_i$ is an example instruction and $\mathbf{z}_i$ is its associated radio-control policy. These demonstrations show how ordered priorities, co-primary objectives, comparisons, and negation should be represented through weights, thresholds, and initialization guidance. For instance, an example can illustrate why ``communication is secondary but must remain reliable'' requires both a smaller communication weight and a nonzero rate threshold, rather than merely assigning communication a low utility weight. The examples affect only the current response; the LLM is not fine-tuned and its parameters remain unchanged.

The two modes otherwise use the same model, schema, response aggregation, validator, and physical solver. This controlled comparison isolates the effect of contextual examples. LLM-ZS is the lower-context default, whereas LLM-ICL is intended to improve calibration for instructions whose meaning depends on combinations, ordering, or negation. The demonstrations are leakage-controlled, with policies, QoS pairs, and wording that are disjoint from the held-out evaluation set.

\subsection{Consensus and Deterministic Validation}
{\color{black}Because generative-model outputs can vary numerically or contain malformed fields, they should not be mapped directly to physical-layer control. To obtain a stable executable policy, the controller generates multiple structured LLM proposals, aggregates numerical fields by their componentwise median, and selects categorical guidance by majority vote before passing the consolidated proposal to the deterministic validator}
\begin{equation}
 \mathbf{z}=\mathcal{V}(\widehat{\mathbf{z}}),
 \label{eq:validator}
\end{equation}
{\color{black}where $\mathcal{V}(\cdot)$ parses the required schema, clips values to their permitted ranges, normalizes the nonnegative priorities so that they sum to one, verifies the QoS thresholds, and maps the initialization field to one of the supported solver modes. Only the validated policy is cached and passed to the radio solver. This validation bounds the generative-AI control interface: the LLM may propose priorities, thresholds, and search guidance, but it cannot directly assign beamforming coefficients, set arbitrary RIS phases, or bypass the physical constraints in~\eqref{eq:main_problem}.}

\subsection{Policy-Conditioned Physical Execution}
{\color{black}The validated policy conditions radio orchestration through feasibility filtering, utility ranking, and search guidance. For each channel state $\mathbf{H}$, the common wireless evaluator constructs the same candidate action set $\cC(\mathbf{H})$ for every language method. Its communication-, sensing-, balanced-, and fairness-oriented modes distribute power and modeled RIS gain differently, and the evaluator returns sum rate, weakest-user rate, sensing SNR, angle CRB, energy, and reconfiguration cost. This shared action space isolates generative-AI mission interpretation from the common deterministic radio solver.}

The compiled thresholds define
\begin{equation}
 \cC_f(\mathbf{z})=\left\{\mathbf{x}\in\cC(\mathbf{H}):R_k\geq R_{\mathrm{th}},\ \forall k,\ \Gamma_s\geq\Gamma_{\mathrm{th}}\right\},
 \label{eq:feasible_set}
\end{equation}
and, when this set is nonempty, the controller selects
\begin{equation}
 \mathbf{x}^{\star}=\arg\max_{\mathbf{x}\in\cC_f(\mathbf{z})}\cF(\mathbf{x};\mathbf{z},\mathbf{H}).
 \label{eq:selection}
\end{equation}
Hence, $R_{\mathrm{th}}$ and $\Gamma_{\mathrm{th}}$ determine which configurations are admissible, $(\lambda_R,\lambda_S,\lambda_F)$ rank those configurations, and $(\rho_0,m_{\mathrm{init}})$ guide the non-convex search. If no candidate satisfies all requirements, the solver first minimizes normalized QoS violation and then compares mission-weighted utilities. The LLM therefore influences the physical solution through $\mathbf{z}$ without directly producing $\{\mathbf{w}_k\}$, $\mathbf{w}_s$, or $\boldsymbol{\phi}$.

\section{Numerical Evaluation}
\subsection{Simulation Setup}
We consider one LEO satellite equipped with an $M=8$ element array, serving $K=3$ representative single-antenna receiver groups and illuminating one sensing target through a fixed ground RIS. The satellite altitude, carrier frequency, and bandwidth are set to $600$~km, $20$~GHz, and $20$~MHz, respectively. User elevation angles are drawn from $25^{\circ}$ to $70^{\circ}$, while the target elevation angle is drawn from $30^{\circ}$ to $65^{\circ}$. Further, we primarily set the RIS size to $N=32$, while $N\in\{8,16,32,64\}$ is considered to examine the effect of surface size. Moreover, the benchmark contains 36 mission prompts, and each prompt is evaluated over 100 paired channel realizations so that all methods experience identical wireless conditions. The main simulation and implementation parameters are summarized in Table~\ref{tab:params}.

The language benchmark contains two complementary sets.\par
\noindent\textit{1) Familiar instructions:} These instructions describe mission types already represented in the classifier taxonomy and therefore test recognition of known operating modes.\par
\noindent\textit{2) Held-out instructions:} These instructions combine priorities, comparisons, and negation in forms absent from both the classifier taxonomy and the in-context demonstrations.

The LLM compiler is implemented using DeepSeek-V4-Flash through DeepInfra, with the sampling temperature fixed to $0.05$, while each mission is queried three times before the numerical fields are aggregated by their componentwise median and the initialization mode is selected by majority vote. LLM-ZS receives no demonstrations, whereas LLM-ICL receives three leakage-controlled mission-policy examples. The prompts, compiled policies, channel realizations, and Monte Carlo seed are frozen for paired evaluation.

Every method selects from the same precomputed set of communication-power fractions and RIS operating modes, with the associated radio metrics computed by the common simulator. The selected beam, power, and RIS configuration is then evaluated using the intended mission weights and QoS thresholds. For method $m$, normalized regret is
\begin{equation}
 \mathcal{R}_m=\frac{\cF_{\mathrm{ref}}(\mathbf{x}^{\star}_{\mathrm{ref}})-\cF_{\mathrm{ref}}(\mathbf{x}_m)}{\cF_{\max}-\cF_{\min}}.
 \label{eq:regret}
\end{equation}
In~\eqref{eq:regret}, $\mathbf{x}^{\star}_{\mathrm{ref}}$ is the best candidate under the intended mission policy and $\mathbf{x}_m$ is selected by method $m$. A smaller $\mathcal{R}_m$ therefore indicates less loss in mission-weighted radio utility. Confidence intervals use a three-level bootstrap over mission families, wording variants, and channel realizations.

\FloatBarrier
\subsection{Performance Analysis}
{\color{black}The proposed generative-AI mission-policy compiler is compared with language, policy, and physical-layer baselines to isolate the value of open-ended semantic adaptation, deterministic validation, and radio execution.} Each test instruction is generated from a stored mission specification containing the intended priorities and minimum communication and sensing requirements. For evaluation only, an \textit{oracle reference} uses this privileged specification to select the best candidate and define the reference utility in \eqref{eq:regret}; because that information is unavailable to a deployed controller, the oracle is not a practical method. The deployable comparison methods are listed below.\par
\noindent\textit{1) TF-IDF classifier:} Maps an instruction to one familiar mission class and applies its fixed policy.\par
\noindent\textit{2) Keyword rules:} Use manually specified lexical conditions to infer the mission.\par
\noindent\textit{3) Demo retrieval:} Copies the policy of the nearest leakage-disjoint demonstration and cannot synthesize a new combination of requirements.\par
\noindent\textit{4) QoS-only:} Uses the correct hard thresholds with equal utility weights, thereby isolating the contribution of weight compilation.\par
\noindent\textit{5) Equal-default:} Applies generic thresholds and equal weights and therefore represents operation without mission adaptation.\par
\noindent Finally, \textit{Random-phase RIS} and \textit{No RIS} isolate the physical-layer contribution of coherent RIS phase control.

\begin{figure}[!t]
\centering
\makebox[\columnwidth][c]{%
\begin{minipage}[b]{0.325\columnwidth}
    \centering
    \includegraphics[width=\linewidth]{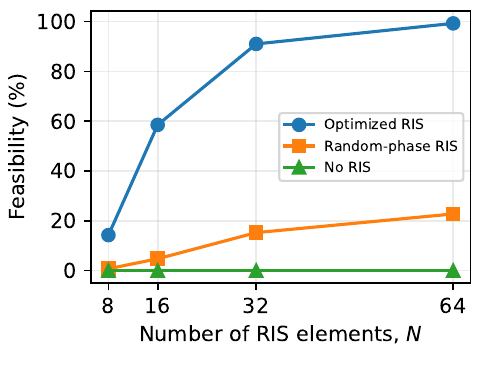}\\[-2pt]
    {\footnotesize (a)}
\end{minipage}%
\begin{minipage}[b]{0.325\columnwidth}
    \centering
    \includegraphics[width=\linewidth]{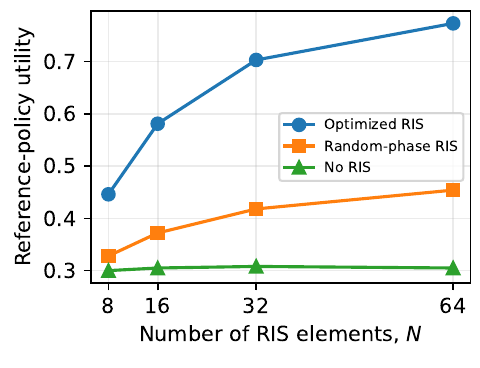}\\[-2pt]
    {\footnotesize (b)}
\end{minipage}%
\begin{minipage}[b]{0.325\columnwidth}
    \centering
    \includegraphics[width=\linewidth]{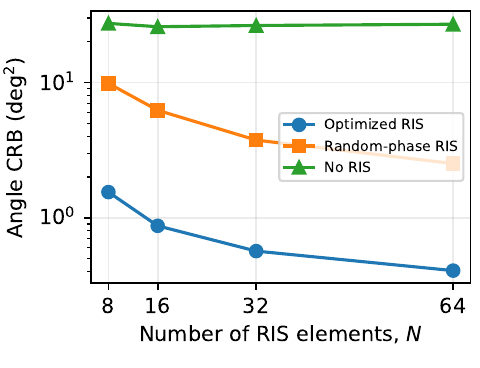}\\[-2pt]
    {\footnotesize (c)}
\end{minipage}}
\caption{RIS-size and phase-optimization performance: (a) joint QoS feasibility; (b) mission-weighted radio utility; (c) angle-estimation CRB versus $N$.}
\label{fig:ris_results}
\end{figure}

\noindent\hspace{1em}\textbf{Physical Benefit of Coherent RIS Optimization:} We first establish whether the deterministic radio solver can physically realize the compiled mission requirements through coherent RIS operation, rather than merely benefiting from a larger reflecting surface. As shown in Fig.~\ref{fig:ris_results}(a), at $N=32$ optimized RIS phases make the joint communication and sensing requirements approximately six times more likely to be satisfied than random phases. Moreover, Fig.~\ref{fig:ris_results}(b) preserves the same ordering under the continuous mission-weighted radio utility, showing that the gain is not limited to configurations that narrowly cross a QoS boundary. Consistently, Fig.~\ref{fig:ris_results}(c) shows that phase optimization lowers the angle-estimation CRB by about $85\%$ at the same surface size. These results provide the physical basis for the mission compiler by showing that flexible priorities are useful only when the radio solver can translate them into coherently optimized communication beams, sensing illumination, and RIS phases.

\begin{figure}[!t]
\centering
\makebox[\columnwidth][c]{%
\subfloat[\footnotesize Policy-weight L1 error.]{%
\begin{minipage}[t]{0.485\columnwidth}\centering
\includegraphics[width=\linewidth]{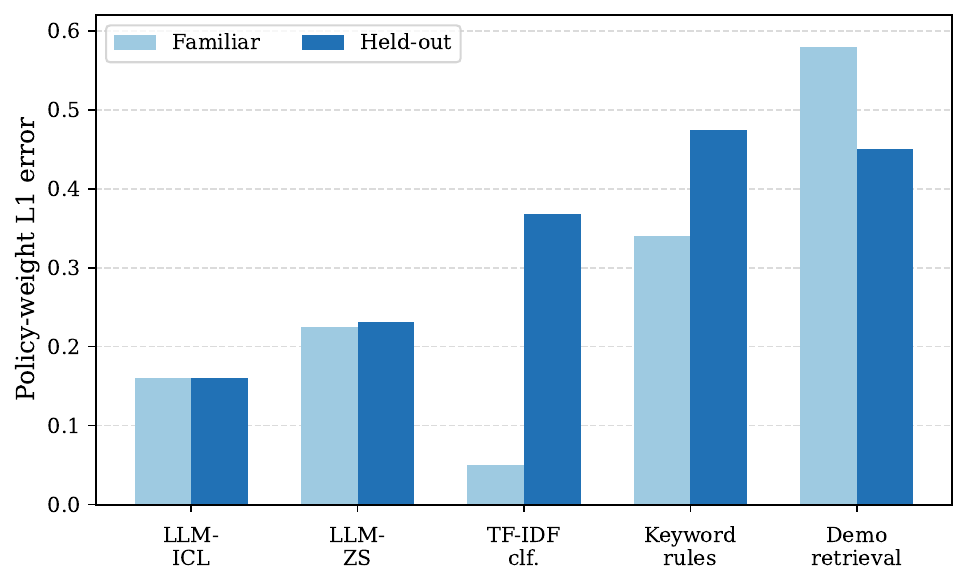}
\end{minipage}}%
\subfloat[\footnotesize Priority-order accuracy.]{%
\begin{minipage}[t]{0.485\columnwidth}\centering
\includegraphics[width=\linewidth]{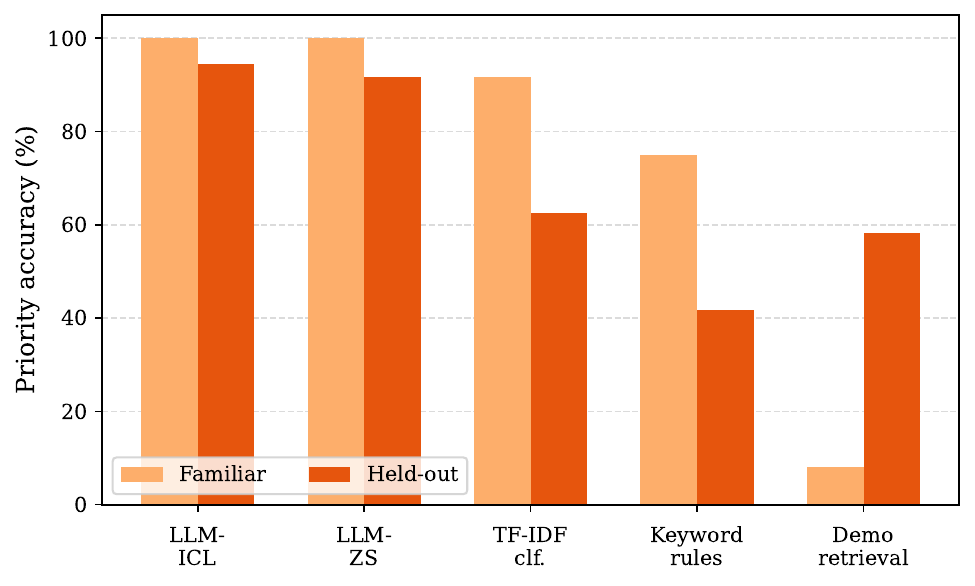}
\end{minipage}}}
\caption{Mission-policy compilation on familiar and held-out instructions.}
\label{fig:semantic_results}
\end{figure}

\textcolor{black}{\textbf{Generative-AI Mission Generalization:}} Having established that optimized RIS control can realize different mission priorities, we next test whether the compiler can recover the requested communication, sensing, and fairness preferences from unfamiliar language. Fig.~\ref{fig:semantic_results}(a) shows that the closed-set classifier is highly accurate for familiar missions, but on held-out missions its weight error rises to $0.3676$, more than twice the $0.1607$ achieved by LLM-ICL, because unseen objective combinations cannot be represented by one predefined class. The priority-order results in Fig.~\ref{fig:semantic_results}(b) reinforce this distinction: both LLM modes remain above $90\%$ on compositional and negated instructions, while the classifier loses nearly 30 percentage points. LLM-ZS is therefore the lower-context choice when identifying the principal objective ordering is sufficient, whereas LLM-ICL is most useful for linguistically difficult missions that require finer numerical calibration. Keyword rules and demo retrieval degrade more sharply because they cannot synthesize new combinations of priorities. QoS-only and Equal-default are omitted here because they do not infer a complete policy from language.

\begin{figure}[!t]
\centering
\includegraphics[width=0.67\columnwidth]{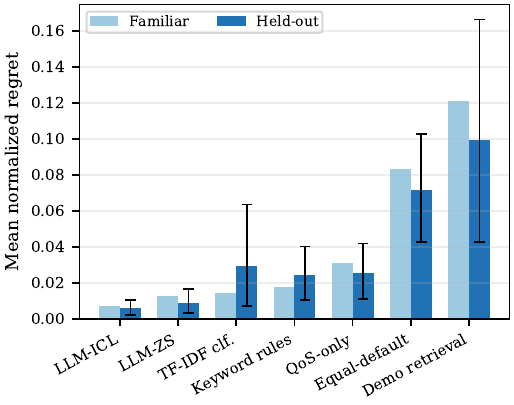}
\caption{\parbox{\columnwidth}{Radio-utility loss caused by mission-policy interpretation errors.}}
\label{fig:regret}
\end{figure}

\textcolor{black}{\textbf{From GenAI Mission Intent to Radio Action:}}
We then determine whether the policy-mapping errors in Fig.~\ref{fig:semantic_results} materially change the selected beamforming, power-allocation, and RIS configuration. Accordingly, Fig.~\ref{fig:regret} shows that both LLM modes remain closest to the oracle reference on held-out missions, whereas fixed and retrieval-based policies incur substantially larger mission-weighted utility loss. The smaller gap between LLM-ICL and LLM-ZS at this stage is informative: both usually recover the mandatory rate and sensing thresholds, and once the feasible set is correct, moderate differences in utility weights often select the same or a nearly equivalent radio configuration. Hence, threshold compilation determines whether a configuration is admissible, while weight calibration mainly refines the choice among admissible configurations. Although LLM-ICL attains the lower mean regret, $0.0060$ versus $0.0089$ for LLM-ZS, the confidence interval of their paired difference includes zero. Zero-shot operation therefore remains the practical default, with ICL reserved for missions requiring finer calibration.

\begin{figure}[!t]
\centering
\includegraphics[width=0.67\columnwidth]{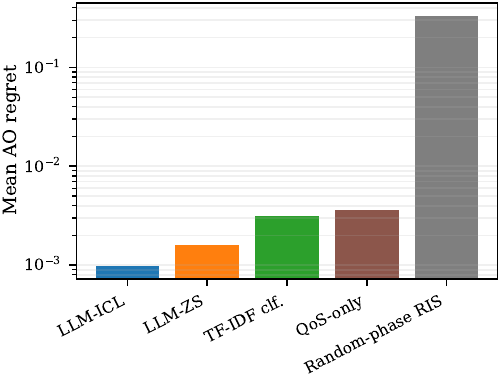}
\caption{Mission-policy performance under explicit beam and RIS optimization.}
\label{fig:ao}
\end{figure}

\textbf{Solver-Level Validation with Explicit Beam and RIS Optimization:} Finally, we test whether the preceding conclusions depend on selecting from the reduced candidate set described in Section~III-D. The reduced validation instead uses an alternating optimization (AO) procedure. For each candidate power split and initialization, the regularized zero-forcing communication beams and target-matched sensing beam are recomputed, after which the unit-modulus RIS coefficients are updated sequentially over a quantized phase grid. These active- and passive-design steps are alternated within the numerical search. Fig.~\ref{fig:ao} shows that the two LLM modes remain the leading mission-aware methods, followed by the classifier and QoS-only, whereas random RIS phases incur more than two orders of magnitude greater regret. The result connects mission interpretation to the underlying non-convex radio problem and confirms that the policy ordering persists under explicit AO of the beams, power split, and RIS phases.

\normalsize\selectfont
\section{Conclusion and Future Work}
{\color{black}
This paper developed a generative-AI-enabled mission-aware radio-orchestration framework for RIS-assisted LEO satellite ISAC systems. The LLM converts natural-language missions into validated utility priorities, hard QoS requirements, and solver guidance, while deterministic optimization retains responsibility for feasibility-critical beamforming, power allocation, and RIS control. Both prompting modes generalize to compositional and negated missions beyond a fixed policy taxonomy. LLM-ICL provides finer semantic calibration, but its lower mean regret ($0.0060$ versus $0.0089$ for LLM-ZS) is not statistically resolved because both modes usually recover the hard QoS thresholds that determine admissibility. Accordingly, LLM-ZS is the low-context default, whereas LLM-ICL is most useful when semantically difficult missions justify additional context. Explicit AO preserves the qualitative ordering when active beams and RIS phases are optimized directly. Overall, the framework shows how generative AI can enhance next-generation network intelligence through adaptable radio objectives and constraints without replacing the deterministic wireless solver. Future work will study online mission revision, uncertainty-aware policy validation, multi-node coordination, Doppler and CSI aging, multiple targets, stronger WMMSE-based optimization~\cite{wmmse}, and measured data.
}

\end{document}